\documentclass[sigconf]{acmart}
\AtBeginDocument{%
  }

\setcopyright{acmlicensed}
\copyrightyear{2018}
\acmYear{2018}
\acmDOI{XXXXXXX.XXXXXXX}
\acmConference[Conference acronym 'XX]{Make sure to enter the correct
  conference title from your rights confirmation email}{June 03--05,
  2018}{Woodstock, NY}
\acmISBN{978-1-4503-XXXX-X/2018/06}

\usepackage[normalem]{ulem}

\usepackage{booktabs}
\usepackage{multirow}
\usepackage{threeparttable}
\usepackage{rotating}
\usepackage{subcaption}
\usepackage{tikz}
\newcommand{\redcirc}[1]{%
  \tikz[baseline=(char.base)]{
    \node[shape=circle, draw=none, fill=red!80,
          inner sep=1.2pt, minimum size=4.5mm,
          text=white, font=\bfseries\footnotesize] (char) {#1};
  }%
}
\begin{document}

\title{SmartPatchLinker: An Open-Source Tool to Linked Changes Detection for Code Review}

\author{Islem Khemissi\footnotemark[1], 
Moataz Chouchen\footnotemark[1], 
Dong Wang\footnotemark[2], 
Raula Gaikovina Kula\footnotemark[3]}
\affiliation{
    \institution{\footnotemark[1]Concordia University, \footnotemark[2]Tianjin University, \footnotemark[3]The University of Osaka\\
    \country{\footnotemark[1]Canada, \footnotemark[2]China, \footnotemark[3]Japan}}
}
\email{khemissi.islem@outlook.com, 
moataz.chouchen@concordia.ca, 
dong_w@tju.edu.cn, 
raula-k@ist.osaka-u.ac.jp}
\date{November 2025}

\renewcommand{\shortauthors}{Khemissi et al.}

\begin{abstract}
In large software ecosystems, semantically related code changes—such as alternative solutions or overlapping modifications—are often discovered only days after submission, leading to duplicated effort and delayed reviews. We present \textbf{SmartPatchLinker}, a browser-based tool that supports the discovery of related patches directly within the code review interface. SmartPatchLinker is implemented as a lightweight Chrome extension with a local inference backend and integrates with Gerrit to retrieve and rank semantically linked changes when a reviewer opens a patch. The tool allows reviewers to configure the search scope, view ranked candidates with confidence indicators, and examine related work without leaving their workflow or relying on server-side installations. We perform both usefulness and usability evaluations to study how SmartPatchLinker can support reviewers during code review. SmartPatchLinker is open source, and its source code, Docker containers, and the replication package used in our evaluation are publicly available on GitHub\footnote{\url{https://github.com/islem-kms/gerrit-chrome-extension}}. A video demonstrating the tool is also available online\footnote{\url{https://drive.google.com/drive/folders/1MCcTj5OSlT7lHVBFMq5m9iatas2joaGb?usp=sharing}}.

.
\end{abstract}

\ccsdesc[500]{Software and its engineering~Software creation and management}

\keywords{Linkage detection, Mining Software Repositories, Code Review}

\maketitle

\section{Introduction}

Modern Code Review (MCR) is a cornerstone of software quality assurance, yet it frequently becomes a bottleneck in fast-paced development cycles \cite{bacchelli2013expectations,badampudi2023modern}. As software projects scale, the increasing volume of submissions fragments development history, making it difficult for developers to maintain awareness of related ongoing work. Prior work defines \emph{linked changes} as independent patches that are semantically related, such as multiple implementations addressing the same bug or feature, but are submitted separately and without explicit references between them \cite{wang2021automatic}. 

When such links are not identified early, developers may unknowingly duplicate effort, resulting in redundant implementations, conflicting changes, and increased review overhead \cite{wang2021automatic}. Identifying semantic relationships between related patches at an early stage is therefore important for efficient code review. Wang et al.~\cite{wang2021automatic} showed that detecting linked patches addressing the same feature request or bug allows reviewers to compare alternative implementations earlier and can speed up the review process. However, their study also reported a persistent notification latency, with a median delay of 3.1 days in large ecosystems such as OpenStack. During this period, developers may continue refining code that is later abandoned in favor of an existing solution, introducing avoidable inefficiencies. Beyond human reviewers, early identification of linked changes is also important for agentic AI systems in software engineering, as it helps build the necessary context to reason about overlapping work, dependencies, and alternative solutions during automated or assisted code review \cite{arabat2024empirical,arabat2025ml}.

Despite increasing interest in automated support for code review \cite{yang2023c3,feng2024machine}, existing techniques remain insufficient to address this latency gap. Early approaches rely primarily on static similarity heuristics, such as TF-IDF-based matching of patch titles and descriptions and file path–based similarity measures \cite{wang2021automatic}. In these methods, lexical and structural signals are combined using fixed or static averaging schemes. While lightweight and easy to deploy, such static aggregation approaches are limited to surface-level overlap and often fail to identify semantically related changes when developers use different terminology, modify different parts of the codebase, or implement alternative solutions to the same problem.

More recently, Large Language Models (LLMs) have shown promise in assisting code review tasks \cite{lu2023llama,pornprasit2024fine,cihan2025evaluating,rao2025overload, trakoolgerntong2025ailinkpreviewer}. However, most LLM-based solutions are deployed as heavyweight server-side services \cite{tang2025just} or chatbot-style assistants \cite{wang2024towards}, which disrupt the reviewer’s in-browser workflow and introduce additional latency, deployment complexity, and privacy concerns. As a result, despite their expressive power, existing LLM-based approaches remain poorly suited for lightweight, real-time assistance during interactive code review.

To address these limitations, we present \textbf{SmartPatchLinker}, a browser-based tool that supports patch linkage detection during code review. SmartPatchLinker integrates directly into the Gerrit interface through a lightweight Chrome extension, allowing reviewers to see related patches when a review is opened. Instead of statically combining TF-IDF and file-based similarity scores, the tool uses semantic text similarity and a learning-based model to combine different similarity signals. SmartPatchLinker runs a local Sentence-BERT (SBERT) backend \cite{reimers2019sentence} to compare change descriptions and does not send code or metadata to external services. Unlike earlier approaches that are applied offline or require server-side installation \cite{wang2021automatic,hirao2019review}, SmartPatchLinker provides feedback directly inside the code review interface.

We evaluate SmartPatchLinker by comparing it against the baseline approaches proposed by Wang \textit{et al.}~\cite{wang2021automatic}, including text-only, file-location-only, and combined variants. Our evaluation shows that SmartPatchLinker consistently outperforms these baselines across multiple large-scale software ecosystems. Using Recall@\(K\), with \(K \in \{1, 2, 4, 6, 8, 10\}\), and Mean Reciprocal Rank (MRR) as ranking metrics, SmartPatchLinker achieves higher recall at low \(K\) values and ranks relevant linkages earlier than lexical and file-based approaches. This behavior is particularly important in interactive code review settings, where reviewers typically inspect only a small number of suggested candidates. These results motivate the practical applicability of SmartPatchLinker as a lightweight, privacy-preserving tool for reducing discovery latency and reviewer effort in modern code review workflows.

\section{SmartPatchLinker}
\begin{figure}[!ht]
    \centering
    \includegraphics[width=\linewidth]{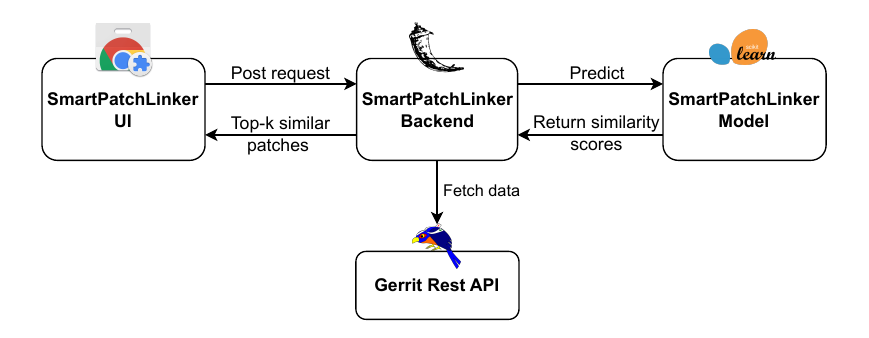}
    \caption{SmartPatchLinker overview.}
    \label{fig:smartpatchlinker_arch}
\end{figure}

Figure \ref{fig:smartpatchlinker_arch} shows an overview of the SmartPatchLinker. 

\subsection{SmartPatchLinker UI}
SmartPatchLinker provides a lightweight, in-context user interface implemented as a Google Chrome extension popup that can be invoked directly from the code review page. When a webpage is loaded, the extension automatically performs \textbf{context detection and injection} by observing the page’s DOM structure to determine whether the current site corresponds to a Gerrit instance\footnote{Specifically, the extension detects the presence of the PolyGerrit \texttt{<gr-app>} root element}. Once verified, it extracts the contextual metadata of the active change, including the \textit{Change-Id}, \textit{project name}, and \textit{patch subject}. This client-side approach avoids administrative deployment on the Gerrit host and enables the tool to operate immediately on any public or private Gerrit instance after installation. The Chrome extension popup acts as the command center of the tool (\redcirc{A}), allowing reviewers to configure the analysis scope by adjusting the lookback window (e.g., 2, 7, 14, or 30 days) and the number of results to retrieve (Top-K), and to explicitly trigger the prediction process using a single action button (\redcirc{B}). Upon execution, the extension forwards the extracted context to the local backend and renders the returned predictions directly within the popup as a ranked list of related patches (\redcirc{C}). Each candidate is annotated with a confidence score and semantic badges (e.g., “95\% match”), providing visual cues that reduce cognitive load and support rapid decision-making, enabling reviewers to quickly identify alternative or overlapping changes without leaving the review interface or disrupting their workflow.
\begin{figure}
    \centering
    \includegraphics[width=0.8\linewidth]{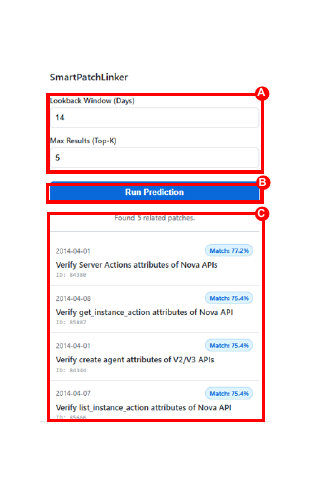}
    \caption{SmartPatchLinker UI overview.}
    \label{fig:smartpatchlinker_UI}
\end{figure}
\subsection{The Backend (Python/Flask API)}
The SmartPatchLinker backend is implemented as a lightweight Python/Flask microservice containerized using Docker, a design choice that decouples computationally intensive processing from the client-side browser extension to ensure responsiveness, scalability, and ease of deployment. The backend includes a dedicated \texttt{GerritClient} module that abstracts the complexities of interacting with the Gerrit REST API, handling authentication and platform-specific security mechanisms, and normalizing the returned JSON responses into a standardized internal representation suitable for downstream processing within the Flask service. On the client side, SmartPatchLinker is implemented as a context-aware Google Chrome extension whose primary role is to detect active Gerrit review sessions and present predictions directly within the review interface, avoiding interruptions to the ongoing code review process and minimizing additional cognitive load for reviewers. To achieve this, the extension performs context detection and injection by observing the page’s DOM structure to identify Gerrit instances—specifically by detecting the PolyGerrit \texttt{<gr-app>} root element—and extracting the \textit{Change-Id}, \textit{project name}, and \textit{patch subject} of the active review. Because this mechanism operates entirely on the client side, the tool can be used immediately on any public or private Gerrit instance without requiring administrative installation on the Gerrit host. Finally, to address browser security constraints such as mixed-content restrictions that prevent HTTPS pages from communicating directly with local HTTP services, SmartPatchLinker employs a secure communication bridge implemented via a background service worker, which proxies requests between the content script and the local Flask backend using the Chrome Runtime API, thereby enabling private, local inference while maintaining browser security guarantees.

\subsection{SmartPatchLinker Model}
The core engine of SmartPatchLinker is built around a feature extraction pipeline that moves beyond traditional lexical matching by incorporating deep semantic understanding through Sentence-BERT. Instead of relying solely on surface-level keyword overlap, the pipeline is designed to capture the underlying intent of code changes, enabling the identification of semantically related patches even when different terminology is used.

\textbf{Candidate Selection.}
Analyzing the full history of a repository for each request is computationally prohibitive. To address this, SmartPatchLinker employs a configurable \emph{temporal window} mechanism that restricts the search space to the active development context. Given a target patch $P$ created at time $t$, the system constructs a candidate set $C = {c_1, c_2, \ldots, c_n}$ such that each candidate $c_i \in C$ satisfies $\mathrm{time}(c_i) \in [t - \delta,, t + \delta]$. In the default configuration, $\delta = 14$ days, which effectively isolates patches submitted within the same development cycle.

\textbf{Feature Engineering.}
For each patch pair $(P, c_i)$, SmartPatchLinker extracts a feature vector $X_i$ composed of three complementary categories of similarity signals.
(1) \textbf{Semantic Text Similarity (SBERT)} captures the intent of the change by generating dense embeddings from the concatenated title and description of each patch using Sentence-BERT, specifically the \texttt{all-MiniLM-L6-v2} model. Semantic similarity is computed as the cosine similarity between embedding vectors $u$ and $v$, defined as $\text{sim}(u, v) = \frac{u \cdot v}{|u| |v|}$. This representation enables the detection of alternative solutions that may be described using different keywords or phrasing.
(2) \textbf{File Path Similarity} extends the file location features proposed by Wang et al.~\cite{wang2021automatic}. File paths are decomposed into segment lists (e.g., \texttt{src/core/utils.py} $\rightarrow$ \texttt{[src, core, utils.py]}), and multiple path-based metrics are computed, including the Longest Common Prefix to capture shared directory structures, the Longest Common Suffix to capture common filenames or extensions, and the Jaccard Index to measure overlap among modified files.
(3) \textbf{Temporal and Structural Meta-Features} complement the semantic and structural signals by incorporating the absolute time difference between patches (in hours) and the difference in the number of modified files (\texttt{delta\_files}), helping the model distinguish immediate follow-up changes from older, unrelated submissions.

\textbf{Prediction and Ranking.}
The extracted feature vectors are passed to a pre-trained Random Forest classifier implemented using \texttt{scikit-learn}, which outputs a probability score representing the likelihood that two patches are semantically linked. Candidate patches are ranked by this score, and the top-$k$ results (with $k = 5$ by default) are returned to the frontend in JSON format for presentation within the user interface.
 

\section{Evaluation}
We compare SmartPatchLinker against three baselines derived from the approach of Wang \textit{et al.}, namely a text-only baseline, a file-location-only baseline, and their combined variant. The comparison is conducted using two complementary ranking metrics: Recall@\(K\), with \(K \in \{1, 2, 4, 6, 8, 10\}\), and Mean Reciprocal Rank (MRR) \cite{nguyen2012duplicate,thongtanunam2015should,ahasanuzzaman2016mining}. Recall@\(K\) evaluates whether at least one relevant linked patch is retrieved within the top-\(K\) ranked candidates, reflecting realistic interactive code review scenarios, while MRR captures how early the first correct linkage appears in the ranking. All approaches are evaluated under identical candidate time windows of \(d \in \{2, 7, 14, 30\}\) days to ensure a fair comparison. It is worth mentioning that we used the data of Wang et al. \cite{wang2021automatic}, based on three large-scale open-source projects: QT (958 linked changes), Android (1,345 linked changes), and OpenStack (9,050 linked changes). 

\subsection{Usefulness evaluation}
Table \ref{tab:mrr_baselines} reports the MRR scores for SmartPatchLinker and the three baseline variants across all evaluated projects.

Higher MRR values indicate that correct linkages are ranked earlier on average. The results show that SmartPatchLinker consistently achieves higher MRR than all baselines, demonstrating its effectiveness at prioritizing relevant linked patches at the top of the ranked list. While the combined baseline improves over the text-only and file-only variants, it remains consistently inferior to SmartPatchLinker, highlighting the benefit of incorporating semantic representations beyond lexical and structural similarity alone.

\begin{figure*}[!ht]
    \centering
    \includegraphics[width=0.9\linewidth]{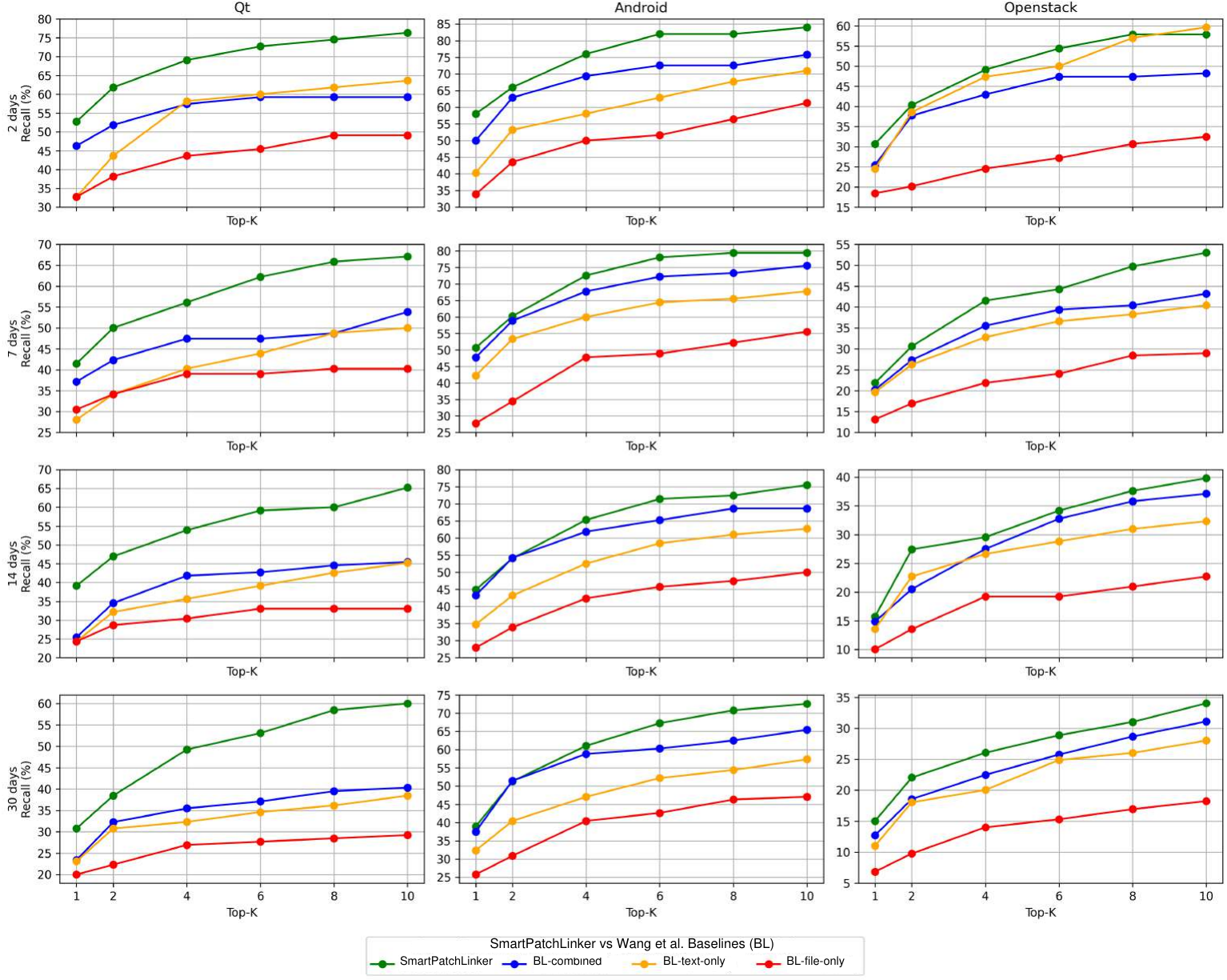}
    \caption{Recall@K scores for SmartPatchLinker compared to Wang et al. \cite{wang2021automatic} baselines.}
    \label{fig:recall_at_k}
\end{figure*}

\begin{table}[!ht]
\centering
\fontsize{6}{6}\selectfont
\tabcolsep=0.03cm
\caption{MRR comparison between SmartPatchLinker and baseline approaches across different time windows.}
\label{tab:mrr_baselines}
\begin{tabular}{p{1.5cm}p{1.25cm}cccc}
\hline
\textbf{Project} & \textbf{Window (days)} &
\textbf{SmartPatchLinker} &
\textbf{BL-combined} &
\textbf{BL-text-only} &
\textbf{BL-file-only} \\
\hline
\multirow{4}{*}{QT}
 & 2  & \textbf{0.60} & 0.52 & 0.45 & 0.57 \\
 & 7  & \textbf{0.49} & 0.42 & 0.32 & \textbf{0.49} \\
 & 14 & \textbf{0.47} & 0.33 & 0.31 & 0.45 \\
 & 30 & 0.39 & 0.30 & 0.27 & \textbf{0.40} \\
\hline
\multirow{4}{*}{Android}
 & 2  & \textbf{0.66} & 0.60 & 0.56 & 0.47 \\
 & 7  & \textbf{0.61} & 0.58 & 0.5 & 0.49 \\
 & 14 & \textbf{0.55} & 0.53 & 0.43 & 0.45 \\
 & 30 & \textbf{0.5} & 0.48 & 0.37 & 0.43 \\
\hline
\multirow{4}{*}{OpenStack}
 & 2  & \textbf{0.4} & 0.35 & 0.23 & 0.26 \\
 & 7  & \textbf{0.31} & 0.28 & 0.15 & 0.17 \\
 & 14 & \textbf{0.22} & \textbf{0.22} & 0.18 & 0.19 \\
 & 30 & \textbf{0.20} & 0.19 & 0.14 & 0.16 \\
\hline
\end{tabular}
\end{table}
Figure \ref{fig:recall_at_k} presents Recall@K results for $K \in \{1, 2, 4, 6, 8, 10\}$ across the Qt, Android, and OpenStack projects.

Each subplot illustrates how recall increases as more candidates are considered, with SmartPatchLinker consistently outperforming all baseline variants across all K values. Importantly, SmartPatchLinker achieves substantially higher recall at low K values (e.g., Recall@1 and Recall@2), which are most relevant for real-time code review settings where reviewers inspect only a small number of suggestions. This trend indicates that SmartPatchLinker is more effective at surfacing correct linkages early, reducing reviewer effort compared to baselines that rely solely on textual or file-based similarity.

\subsection{Usability evaluation}
We assess the usability of \textbf{SmartPatchLinker} by examining the time required for reviewers to obtain actionable results during a typical code review session. Across our evaluation scenarios, SmartPatchLinker completes the end-to-end interaction—starting from opening a review, configuring the analysis parameters, triggering the prediction, and inspecting the returned linked patches—in under two minutes for the majority of cases. This includes both the local inference time and the user interaction overhead. The short execution time ensures that SmartPatchLinker integrates smoothly into existing review workflows without introducing noticeable delays or disrupting the reviewer’s cognitive flow. These findings indicate that the tool is sufficiently lightweight for real-time, interactive use in modern code review environments.

\section{Conclusion \& Future Work}
In this paper, we introduced \textbf{SmartPatchLinker}, a browser-based assistant that brings semantic patch linkage detection directly into the modern code review process. By separating the user interface (Chrome extension) from the core analysis engine (local SBERT-based inference), SmartPatchLinker provides in-context feedback while remaining lightweight and privacy-aware. As future work, we plan to extend SmartPatchLinker beyond Gerrit to support GitHub Pull Requests and GitLab Merge Requests, enabling broader adoption across development platforms. We also plan to explore tighter integration with LLM-based and agentic systems, for example, by summarising detected dependencies between related patches, highlighting alternative solutions, or providing short explanations of how changes overlap. These directions aim to further support both human reviewers and automated agents in building context and making informed review decisions.

\appendix

\section{Use case \& Walkthrough}
Before diving into the details, watch our \href{https://drive.google.com/drive/folders/1MCcTj5OSlT7lHVBFMq5m9iatas2joaGb?usp=sharing}{short demo video} to see how SmartPatchLinker identifies links in real-time on the OpenStack Gerrit interface.

\subsection{Scenario}
Consider a maintainer for the OpenStack project reviewing a newly submitted patch with a given title. The patch might appear to be technically sound, but the maintainer has little to no recollection of a similar attempt made by another contributor that might have happened days or weeks ago. Relying on Gerrit's native "Related Changes" tab yields no results, as the previous attempt was on a different branch and shares no Git history. Normally, the reviewer would have to rely on manual keyword searches or simply approve the patch, potentially introducing technical debt or redundancy.

\subsection{Workflow}
\begin{figure}
    \centering
    \includegraphics[width=1\linewidth]{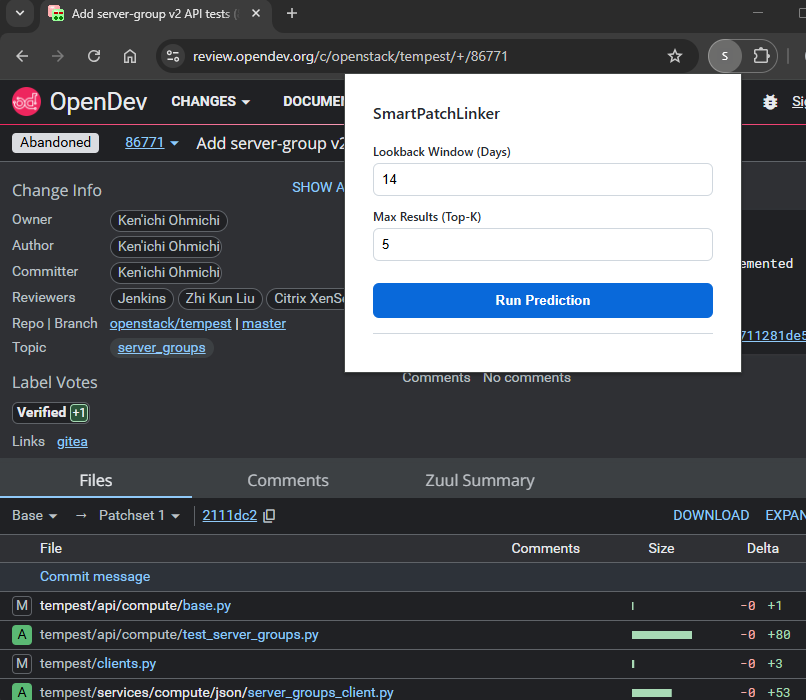}
    \caption{Configuration: Time Window and Top-K}
    \label{fig:extension_configuration}
\end{figure}

\begin{figure}
    \centering
    \includegraphics[width=1\linewidth]{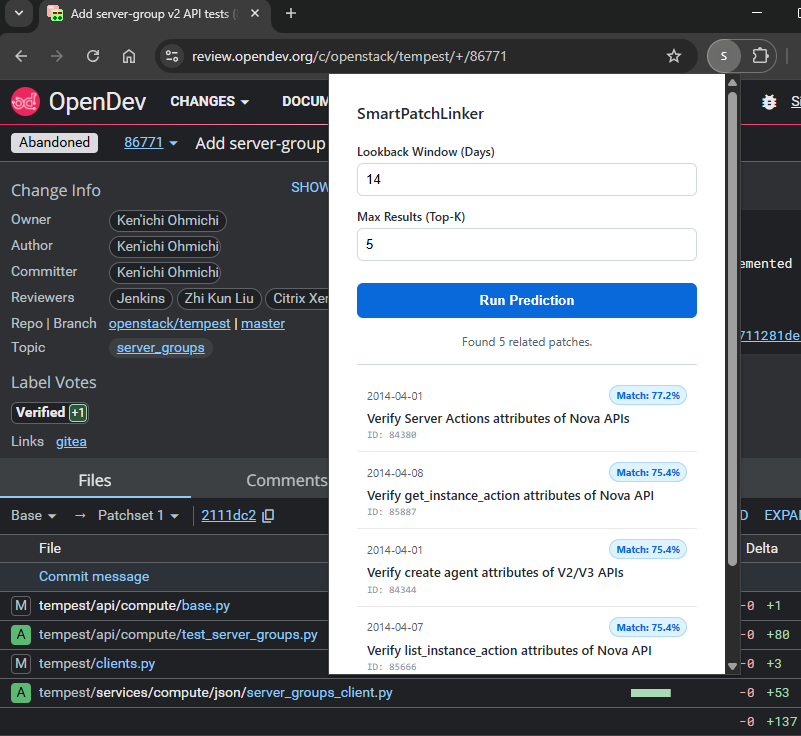}
    \caption{Top-K Results}
    \label{fig:results}
\end{figure}

Instead of manually searching, the reviewer can use SmartPatchLinker:
\begin{enumerate}
    \item \textbf{Context Recognition}: When navigating to the patch page, the extension detects the URL and extracts the \texttt{Change-Id} and \texttt{Project}.
    \item \textbf{Configuration}: The reviewer clicks on the extension icon to open the popup interface (\autoref{fig:extension_configuration}). The \textbf{Time Window} and \textbf{Top-K} values can be adjusted according to the reviewer's needs. Then, the prediction process can be triggered via the button \textit{"Run Predictions"}
    \item \textbf{Semantic Discovery}: The extension forwards the request to the local middleware where the candidates are extracted based on the \textit{Time Window} specified from the previous step. The SBERT model analyzes the semantic intent of the current patch and the candidates.
    \item \textbf{Result Analysis}: The tool returns the $k$ number of predictions with their respective confidence in percentage (\autoref{fig:results}).
    \item \textbf{Resolution}: The reviewer can then check the results to confirm any linkage to the current change. If any change is confirmed, the reviewer can then add a comment with the URL of the linked change that can be seen by the developer or any other user that can access the patch.
\end{enumerate}


\bibliographystyle{ACM-Reference-Format}

\bibliography{references}

\end{document}